\documentclass[a4paper]{jpconf}
\usepackage{graphicx}
\usepackage{subfigure}
\usepackage{caption}
\usepackage{url}
\usepackage{hyperref}
\usepackage{lineno}

\begin{document}
\title{The MicroBooNE continuous readout stream for detection of supernova neutrinos}

\author{J I Crespo-Anad\'on, on behalf of the MicroBooNE collaboration}

\address{Columbia University, New York, NY, 10027, USA}

\ead{jcrespo@nevis.columbia.edu}

\begin{abstract}
Since the original detection of core-collapse supernova neutrinos in 1987, all large neutrino experiments seek to detect the neutrinos from the next nearby supernova. Among them, liquid argon time projection chambers (LArTPCs) offer a unique sensitivity to the electron neutrino flux of a supernova. However, the low energy of these events (scale of MeVs), and the fact that all large (multi-tonne) LArTPCs operating at the moment are located near the Earth's surface, and are therefore subject to an intense cosmic ray flux, makes triggering on the supernova neutrinos very challenging. Instead, MicroBooNE has pioneered a novel approach for detecting supernova neutrinos based on a continuous readout stream and a delayed trigger generated by other neutrino detectors (the Supernova Early Warning System, or SNEWS). MicroBooNE's data is stored temporarily for a few days, awaiting a SNEWS alert to prompt the permanent recording of the data. In order to cope with the large data rates produced by the continuous readout of the TPC and the PMT systems of MicroBooNE, FPGA-based zero-suppression algorithms have been developed.
\end{abstract}

\section{The MicroBooNE detector}
The MicroBooNE detector is a liquid argon time projection chamber located in the Booster Neutrino Beam at Fermilab, and started beam data taking in October 2015. The experiment goals are investigating the excess of electron-like events observed by MiniBooNE~\cite{Aguilar-Arevalo:2018gpe}, performing high-precision measurements of cross-section of $\nu_{\mu}$ and $\nu_{\rm e}$ on argon, developing further the LArTPC technology and searching for astroparticles and exotic physics using the LArTPC capabilities, both on and off-beam.

MicroBooNE has an active mass of 90 tonnes, with a maximum drift length of 2.5 m, corresponding to a drift time of 2.3 ms. Three wire planes with a 3 mm wire pitch are used to detect the ionization charge left by interacting particles: 2 induction planes with 2400 wires each at $\pm 60^\circ$ from the vertical, and a collection plane with 3256 vertical wires. In addition, 32 PMTs with TPB-coated acrylic plates mounted behind the wire planes are used to detect the scintillation light and provide a trigger signal for the detector. The detector is further described in Ref.~\cite{Acciarri:2016smi}.

\section{Detection of core-collapse supernova neutrinos using a delayed trigger}
The neutrinos emitted by a core-collapse supernova are expected to occur in a short interval spanning tens of seconds, and with energies ranging in the tens of MeV. The expected number of events in the MicroBooNE detector for a canonical supernova at 10 kpc is of the order of ten events (scaling from the expected number of events in the DUNE detector~\cite{Acciarri:2015uup}), detected through the charged-current interaction $\nu_{\rm e} + \rm{{}^{40}Ar} \to e^- + \rm{{}^{40}K^*}$, with a threshold of $\sim 5$ MeV. This channel provides a unique sensitivity to the $\nu_{\rm e}$ flux, which is complementary to the predominant $\overline{\nu}_{\rm e}$ sensitivity of the water Cherenkov and liquid scintillator detectors.

Due to the near-surface location, the small number of neutrino interactions expected and their low-energy, MicroBooNE cannot effectively trigger on these events with the PMT system. Instead, MicroBooNE relies on an external trigger given by the SNEWS Alert Mailing List~\cite{SNEWSMailingList, Antonioli:2004zb}. This alert is initially targeted to the astronomical community as the light counterpart of the supernova is expected to arrive within a few hours of the neutrino detection, but it is repurposed by MicroBooNE as a delayed trigger for data acquisition. In order to do this, the MicroBooNE detector is continuously read out and the data is stored in disks in the DAQ servers for a few hours awaiting the SNEWS alert to transfer it to permanent storage or be deleted otherwise.

In addition to the detection of core-collapse supernova neutrinos, the continuous readout enables the acquisition of non-triggered data for the continuous monitoring of the detector, or searches for non-beam events such as nucleon decay. While the size of MicroBooNE prevents competitive searches for these processes, the data can be used to prototype future analyses in larger detectors or study backgrounds.

\section{The MicroBooNE triggered and continuous readouts}
\begin{figure}
\begin{center}
\includegraphics[width=0.98\textwidth]{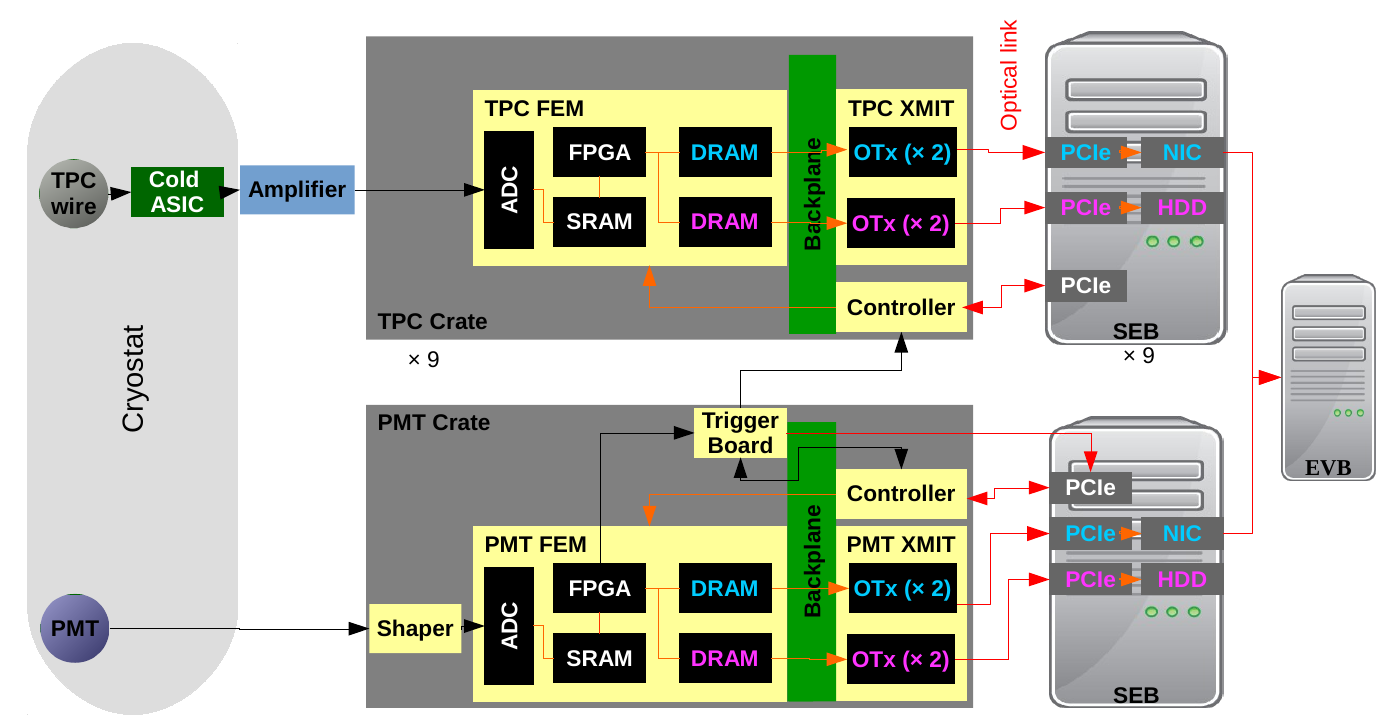}
\end{center}
\caption{Diagram of the MicroBooNE readout. The triggered stream components are highlighted in blue text and the continuous readout stream components are highlighted in magenta.}
\label{fig:ReadoutElectronics}
\end{figure}

The MicroBooNE readout electronics are shown in Fig.~\ref{fig:ReadoutElectronics}. The signal in each TPC wire is preamplified and shaped by an ASIC immersed in the liquid argon and subsequently extracted from the cryostat through a feedthrough. The signal is further amplified by warm electronics in order to prepare it for transmission 
to the readout electronics using shielded twisted-pair cables. The signal of 64 wires is received by the Front-End Module (FEM), where it is digitized by a commercial 12-bit ADC 
at 16 MSps, later downsampled to 2 MSps and written to a $1~\rm{M} \times 36~\rm{bit}$ $128~\rm{MHz}$ static RAM (SRAM) in time order. The SRAM is configured as a ring buffer, storing $12.8~\rm{ms}$ of data split into 8 frames ($1.6~\rm{ms/frame}$). From here, an FPGA (Altera Stratix III) reads out the TPC signal in channel order and splits it into two streams. The triggered stream is only read out upon trigger, and has been described in detail in Ref.~\cite{Acciarri:2016smi}. The supernova stream is continuously read out. The data of each FEM is sent to a transmitter board (XMIT) through a backplane with bandwidth up to $512~\rm{MB/s}$. The backplane dataway is shared between both streams, with the triggered stream given priority over the continuous stream using a token-passing scheme. The FEM has a dynamic RAM (DRAM) for each stream to store the data awaiting its turn to be transferred. The transmitter board has 4 optical transceivers, each rated to $3.125~\rm{Gbps}$. Two are used for the triggered stream, and two are used for the supernova stream. Finally, the data is read out by custom PCIe $\times 4$ cards connected to the PCIe bus of the Sub-Event Buffer (SEB) DAQ server. The triggered data is sent to the Event Builder (EVB) DAQ server using a network interface card (NIC), while the continuous stream is written to the local hard disk drive (HDD). The bottleneck of the continuous readout chain is the HDD writing, which is assumed to be $50~\rm{MB/s}$ conservatively.

The readout of the TPC is approximately equally distributed among 9 SEBs. Each SEB is equipped with a local HDD of 13 TB, upgraded to 15 TB in the summer of 2018. A process monitors the occupancy of the disk, and whenever it goes above $80\%$, deletes the oldest run(s) until the occupancy is below $70\%$. Writing at $50~\rm{MB/s}$, this gives  $> 48~\rm{h}$ to respond to a SNEWS alert.

A similar stream exists for the PMT readout, as described in Ref.~\cite{Kaleko:2013eda}.

\section{Compression algorithms for the continuous TPC readout stream}
Without compression, the MicroBooNE TPC readout generates $\sim 33~\rm{GB/s}$.
Each SEB would need to read $\sim 3.7~\rm{GB/s}$. 
In order to bring down the data rates to the $50~\rm{MB/s}$ target, a series of compression algorithms are needed to reduce the data volume by a factor of $\sim 80$. The lossless Huffman compression that is also used for the triggered stream is not enough to achieve this goal and an additional lossy compression is required, which is implemented in the FEM FPGA. 

\begin{figure}
\begin{center}
\includegraphics[width=0.5\textwidth]{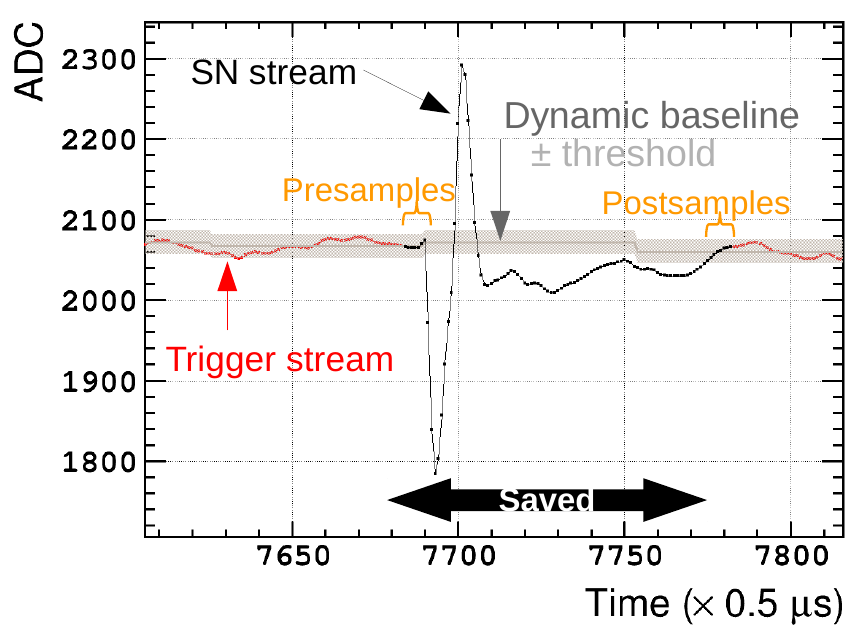}
\end{center}
\caption{Example of data from the test stand at Nevis Laboratories showing a zero-suppressed waveform in the continuous readout stream (black), superimposed on the same waveform from the triggered stream (red). An emulation of the dynamic baseline used by the FPGA is shown as a gray line, with the threshold shown as a light gray band. Only the samples out of this band are saved, plus a number of samples preceding them (presamples) and following them (postsamples).}
\label{fig:ZeroSuppression}
\end{figure}

Several lossy compression algorithms have been tested. All of them rely on zero suppression by recording only the deviations of the waveforms 
that pass an amplitude threshold with respect to a baseline. For each group of contiguous samples passing the threshold, a number of samples preceding the first sample and trailing the last are also recorded. 
This allows to capture the portion of the signal under the threshold, determine the local baseline (since the baseline used by the electronics is not written to the output data) and help the deconvolution of signals. An example of a zero-suppressed waveform is shown in Fig.~\ref{fig:ZeroSuppression}. The compression algorithms differ in how the baselines and thresholds are implemented.

\paragraph{Dynamic baseline} The signal is split into blocks of 64 samples each, and the mean and a truncated variance are computed per block. The result for each block is compared to the preceding and following blocks, and if within tolerances (configurable per FEM), the mean value of the central block is used as baseline.
\paragraph{Static baseline} The baseline for each channel is loaded at the beginning of the run, and kept fixed during the run.

\paragraph{Plane-wide thresholds} All the channels belonging to the same TPC plane are assigned the same threshold (configurable per FEM).
\paragraph{Channel-wise thresholds} Each channel is assigned an individualized threshold. \newline

\begin{figure}[b]
\centering
        \subfigure[]{
            \includegraphics[width=0.98\linewidth]{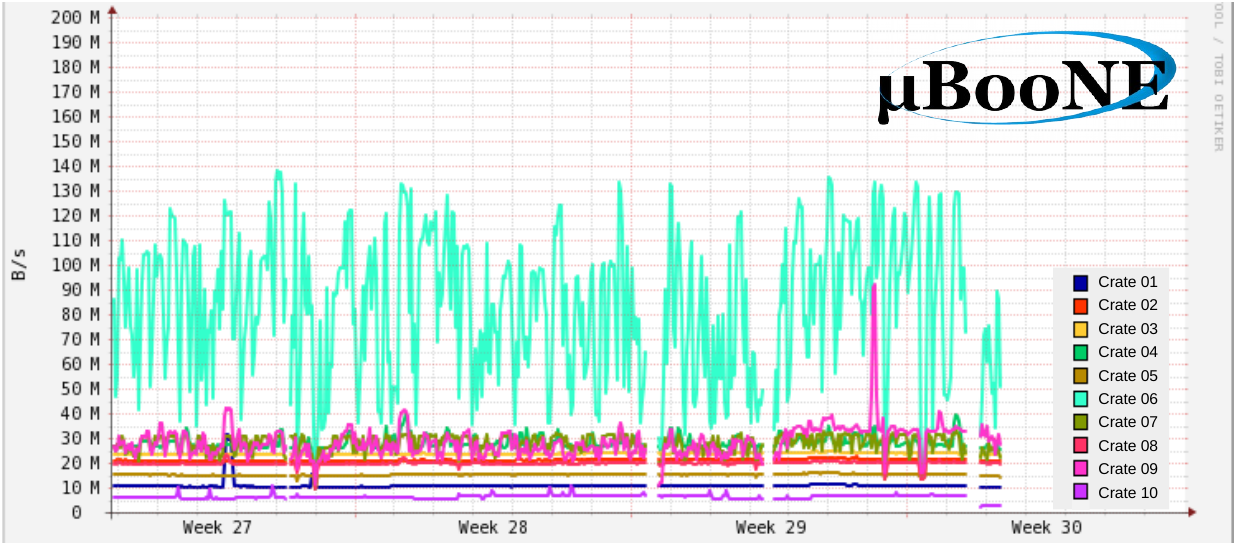}
            \label{fig:DataRatesDynamicBaselinePlaneThresholds}
        }

\end{figure}

\begin{figure}[hp]
\centering
        \subfigure[]{
            \includegraphics[width=0.98\linewidth]{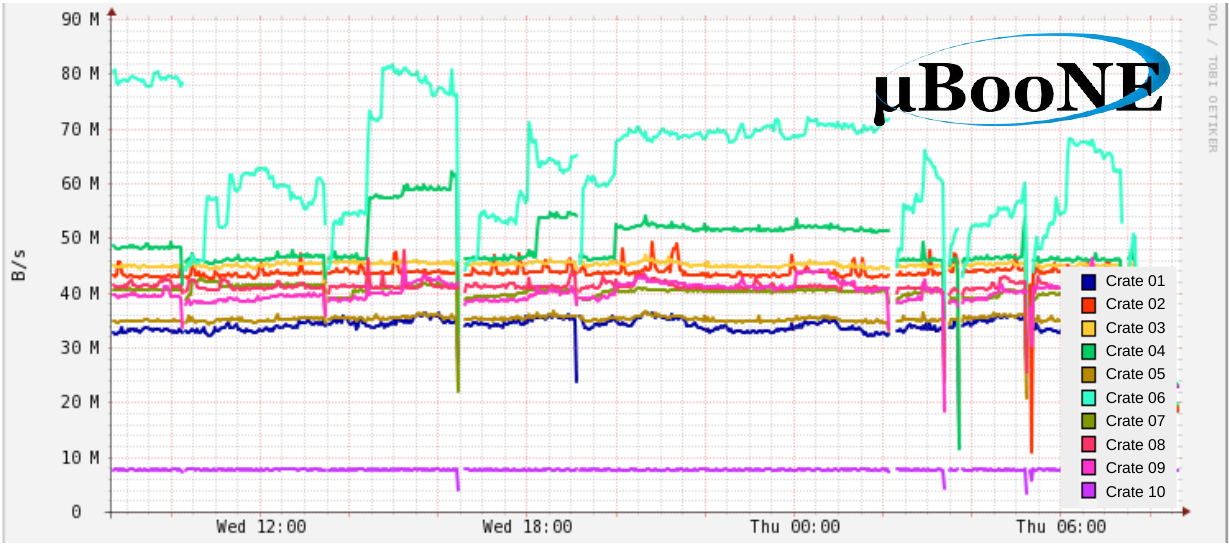}
            \label{fig:DataRatesDynamicBaselineChannelThresholds}
        }
        \subfigure[]{
            \includegraphics[width=0.98\linewidth]{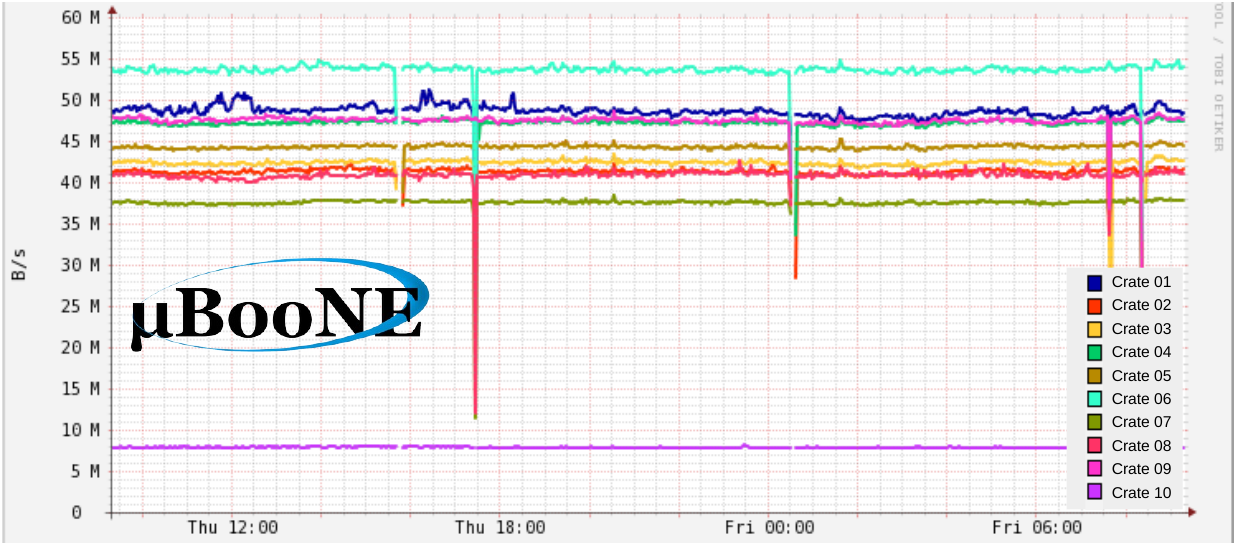}
            \label{fig:DataRatesStaticBaselineChannelThresholds}
        }
	  \caption{Disk writing rates from the continuous readout stream measured by the Ganglia monitoring system. Each trace shows the rate from the DAQ server connected to each crate. Crates 01 -- 09 read out the TPC wires, and crate 10 reads out the PMTs.
	  \subref{fig:DataRatesDynamicBaselinePlaneThresholds} corresponds to a compression algorithm that uses a dynamic baseline estimation for each channel and a common threshold for all the channels belonging to the same TPC plane.
	  \subref{fig:DataRatesDynamicBaselineChannelThresholds} corresponds to a compression algorithm that uses a dynamic baseline and channel-wise thresholds.
	  \subref{fig:DataRatesStaticBaselineChannelThresholds} corresponds to a compression algorithm that uses static baselines and channel-wise thresholds.}
	\label{fig:DataRates}
\end{figure}

Fig.~\ref{fig:DataRates} shows the data rates resulting from the compression algorithms tested. Fig.~\ref{fig:DataRatesDynamicBaselinePlaneThresholds} shows the results using dynamic baselines and plane-wide thresholds. The thresholds were determined by analyzing jointly the waveforms from all the channels from previous triggered stream runs, and searching for a valley between the noise and signal peaks in the ADC distribution. This results in data rates that are well below the $50~\rm{MB/s}$ target except for crate 06, which is connected to a group of especially noisy channels. In fact, crate 06 experiences large variations in data rates since the noisy channels interfere with the establishment of a proper baseline.

Fig.~\ref{fig:DataRatesDynamicBaselineChannelThresholds} shows the data rates for a compression algorithm that uses dynamic baselines but channel-wise thresholds. The thresholds were determined by analyzing previous data from the triggered stream, and finding the symmetric thresholds which encompass $98.5\%$ of the ADC distribution for each channel, which result in data rates closer to the $50~\rm{MB/s}$ target. This resulted in smaller thresholds overall, increasing the sensitivity to smaller signals, and the data rates increased as expected, but the large variation in crate 06 remained.

Fig.~\ref{fig:DataRatesStaticBaselineChannelThresholds} shows the data rates of the compression algorithm that uses static baselines and channel-wide thresholds. It improves on the stability of the data rates compared to the previous algorithm versions by removing the dynamic estimation of the baseline, which is too sensitive to noisy channels. This is the current compression algorithm in use in the MicroBooNE continuous stream.


\begin{figure}
\centering
            \includegraphics[clip, trim= 2cm 1.5cm 1cm 2cm,width=0.55\linewidth]{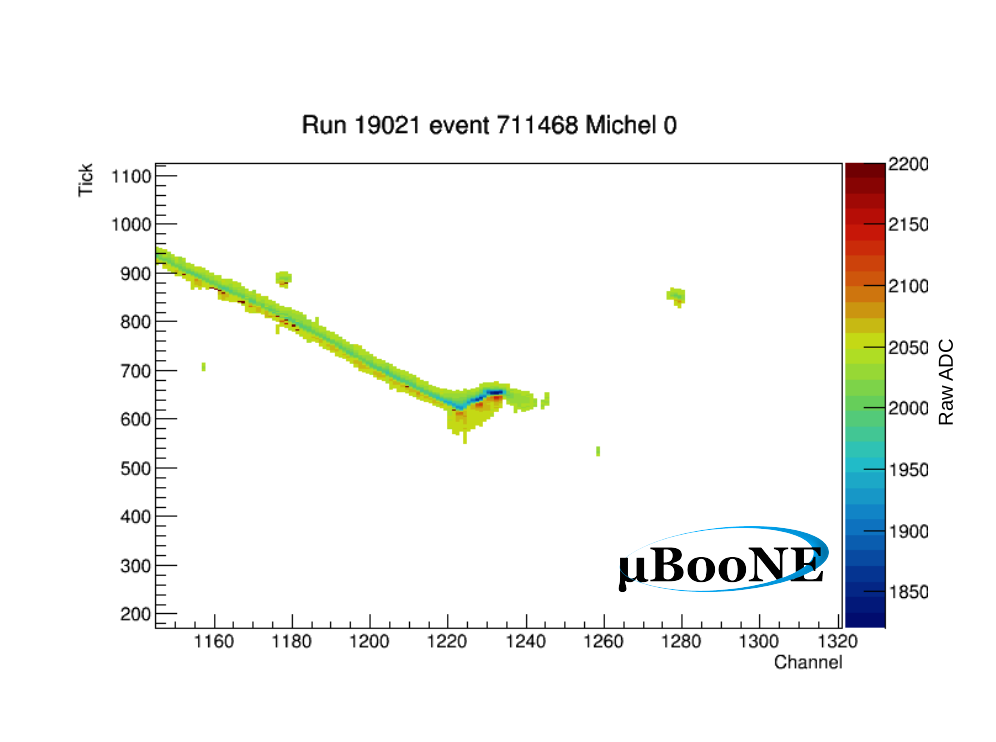}
            \includegraphics[clip, trim= 2cm 1.5cm 1cm 2cm,width=0.55\linewidth]{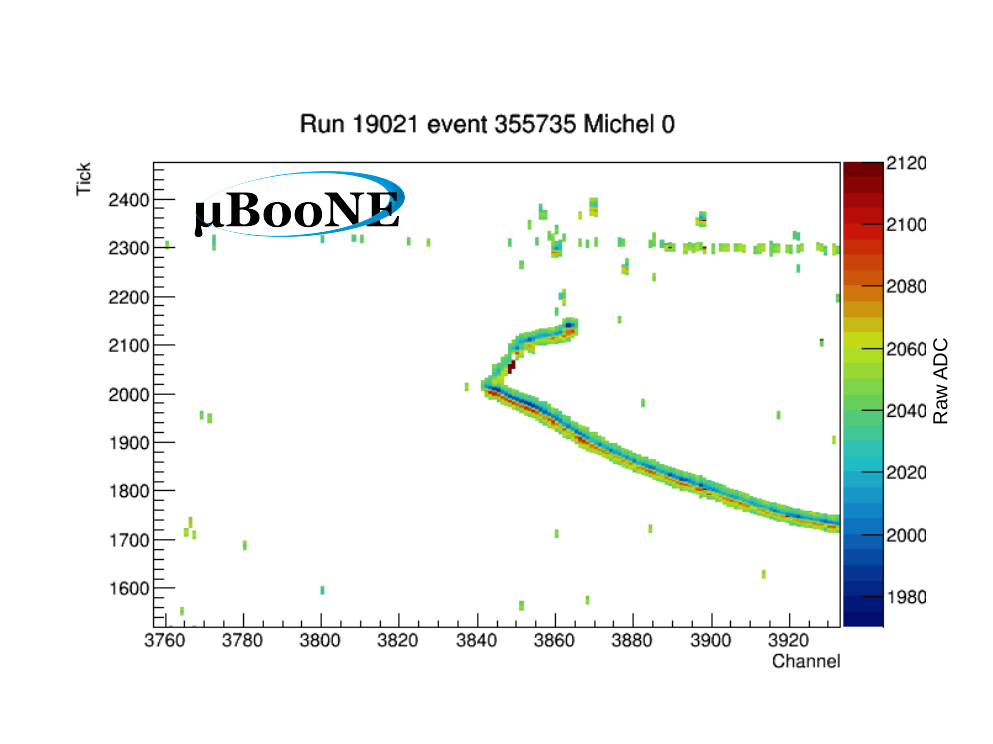}
            \includegraphics[clip, trim= 2cm 1.5cm 1cm 2cm,width=0.55\linewidth]{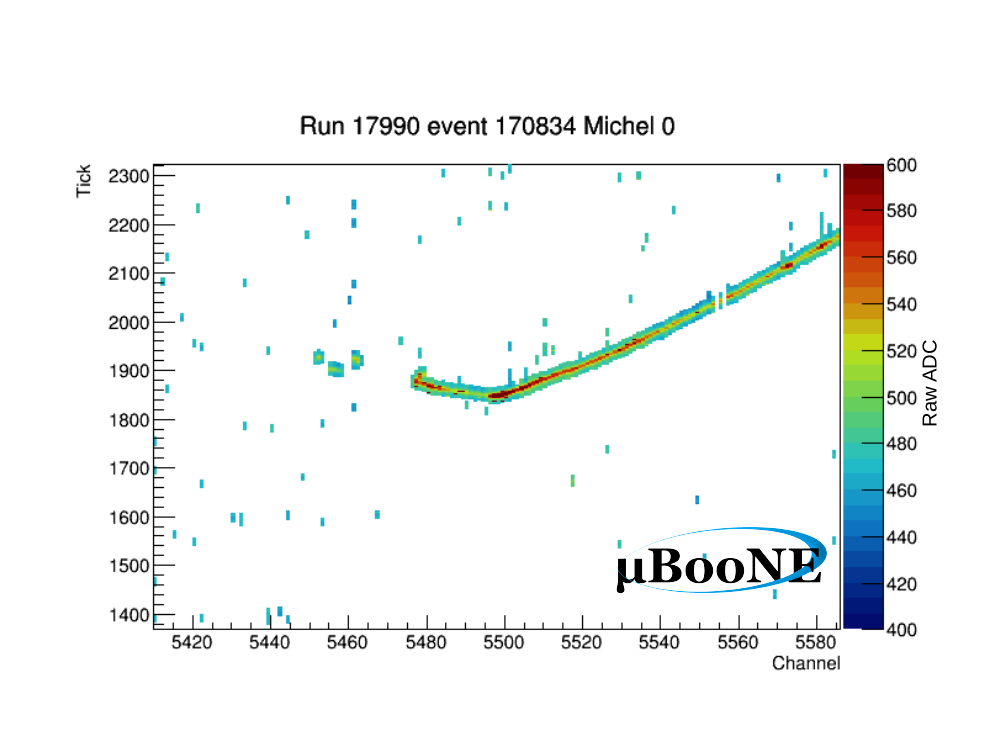}
	  \caption{Michel electron candidates found in the continuous readout stream on the three TPC planes: first induction plane (top), second induction plane (middle) and collection plane (bottom). The horizontal axis shows the TPC wire number and the vertical axis shows the sample number (tick) in the 2 MHz clock. The color scale shows the uncalibrated ADC scale. The white background corresponds to the data which has been zero-suppressed.}
	\label{fig:MichelCandidates}
\end{figure}

In order to demonstrate the sensitivity of the continuous readout stream, Michel electrons are being used as a proxy for supernova neutrino interactions in the LArTPC since the spectrum of these electrons overlaps with the spectrum of those from core-collapse supernova neutrino interactions~\cite{Acciarri:2017sjy}. The analysis is in progress but multiple Michel electron candidates have already been found on the three TPC planes (see Fig.~\ref{fig:MichelCandidates}) using the same automated reconstruction and selection described in Ref.~\cite{Acciarri:2017sjy}, confirming the capability of the continuous readout stream to detect low energy interactions.

\section{Conclusion}
MicroBooNE has commissioned a new continuous readout stream for the detection of core-collapse supernova neutrinos using the SNEWS alert as delayed trigger. In order to cope with the large data rates, a compression algorithm based on zero suppression is applied online by an FPGA. The best performance is found with an algorithm that uses static baselines and individualized thresholds for each readout channel, reaching the $\sim 80$ compression factor target. The observation of Michel electron candidates on the three TPC planes opens the way to a potential detection of supernova neutrinos in MicroBooNE. Furthermore, the zero-suppressed TPC waveforms could be fed into a trigger algorithm to generate detector triggers based on the TPC information, which is of interest for the future DUNE experiment.


\section*{References}
\begin{thebibliography}{9}
\bibitem{Aguilar-Arevalo:2018gpe} 
  A.~A.~Aguilar-Arevalo {\it et al.} [MiniBooNE Collaboration],
  Phys.\ Rev.\ Lett.\  {\bf 121}, no. 22, 221801 (2018)
  [arXiv:1805.12028 [hep-ex]].

\bibitem{Acciarri:2016smi} 
  R.~Acciarri {\it et al.} [MicroBooNE Collaboration],
  JINST {\bf 12}, no. 02, P02017 (2017)
  [arXiv:1612.05824 [physics.ins-det]].

\bibitem{Acciarri:2015uup} 
  R.~Acciarri {\it et al.} [DUNE Collaboration],
  arXiv:1512.06148 [physics.ins-det].
  
\bibitem{SNEWSMailingList} 
  \textit{Getting a SNEWS alert}, \href{https://snews.bnl.gov/alert.html}{https://snews.bnl.gov/alert.html}. Accessed on March 3, 2019.

\bibitem{Antonioli:2004zb} 
  P.~Antonioli {\it et al.},
  New J.\ Phys.\  {\bf 6}, 114 (2004)
  [astro-ph/0406214].
  
\bibitem{Kaleko:2013eda} 
  D.~Kaleko,
  JINST {\bf 8}, C09009 (2013)
  [arXiv:1308.3446 [physics.ins-det]].

\bibitem{Acciarri:2017sjy} 
  R.~Acciarri {\it et al.} [MicroBooNE Collaboration],
  JINST {\bf 12}, no. 09, P09014 (2017)
  [arXiv:1704.02927 [physics.ins-det]].
  
\end{thebibliography}
\end{document}